# Beyond Pipelines: A Fundamental Study on the Rise of Generative-Retrieval Architectures in Web Research

Amirreza Abbasi
Department of Computer Science and Information Technology
Institute for Advanced Studies in Basic Sciences
Zanjan, Iran
a.abbasi@iasbs.ac.ir

Mohsen Hooshmand
Department of Computer Science and Information Technology
Institute for Advanced Studies in Basic Sciences
Zanjan, Iran
mohsen.hooshmand@iasbs.ac.ir

*Abstract*— Web research and practices have evolved significantly over time, offering users diverse and accessible solutions across a wide range of tasks. While advanced concepts such as Web 4.0 have emerged from mature technologies, the introduction of large language models (LLMs) has profoundly influenced both the field and its applications. This wave of LLMs has permeated science and technology so deeply that no area remains untouched. Consequently, LLMs are reshaping web research and development, transforming traditional pipelines into generative solutions for tasks like information retrieval, question answering, recommendation systems, and web analytics. They have also enabled new applications such as web-based summarization and educational tools. This survey explores recent advances in the impact of LLMs-particularly through the use of retrieval-augmented generation (RAG)-on web research and industry. It discusses key developments, open challenges, and future directions for enhancing web solutions with LLMs.

## I. INTRODUCTION

The emergence of the Internet fundamentally transformed modern society. Today, a few human activities ranging from education and politics to commerce and scientific research remain unaffected by the Internet and the World Wide Web. As a result, nearly every discipline has experienced structural changes driven by its increasing reliance on web-based infrastructures and digital interactions [1].

From a computer science perspective, one of the primary challenges in developing web-based systems broadly referred to as web research has been the management and utilization of large-scale, heterogeneous data. Prior to the rapid advancement of artificial intelligence (AI), web systems largely relied on classical, modular, and task-specific algorithms to address challenges such as information retrieval, recommendation, and content analysis. However, the recent acceleration in AI development, particularly with the rise of LLMs, has introduced a paradigm shift. LLMs blur the traditional boundary between retrieving information and generating content [2][3], while simultaneously enabling semantic reasoning, contextual understanding, content synthesis, decision-making, and interactive agent capabilities. In this sense, the development of LLMs has shifted web research from pipeline-based architectures toward integrated generative–retrieval ecosystems [4], [5].

Web research is a vast interdisciplinary domain spanning areas from cybersecurity to computational social science. Although LLMs influence many of these subfields, this survey focuses on foundational pillars of web research, including information retrieval, recommender systems, conversational agents, web analytics, and web-based education platforms. Particular emphasis is placed on RAG architectures as a central mechanism for enhancing LLM capabilities in web applications by integrating parametric and external knowledge sources [6]. Accordingly, this paper provides a cross-domain synthesis of LLM integration across core web research areas. While numerous specialized surveys examine individual components of web-based systems, they are often confined to isolated domains. In contrast, this work aggregates as shown in Fig. 1 and systematizes these perspectives to offer a unified starting point for researchers seeking a broader understanding of LLM-driven web systems. We analyze architectural patterns across domains, discuss shared technical and ethical challenges, and outline emerging research directions that shape the future of intelligent web infrastructures.

## II. BACKGROUND

LLMs are large-scale Transformer-based autoregressive neural networks trained on extensive web-scale corpora. By optimizing next-token prediction objectives over diverse textual data, these models acquire broad linguistic competence and implicit world knowledge encoded within their parameters. This parametric knowledge enables LLMs to perform a wide range of tasks including question answering, summarization, reasoning, and dialogue without task-specific architectural modifications. Their strength lies in contextual language understanding and generative flexibility, allowing them to synthesize coherent and semantically meaningful outputs across heterogeneous domains [1].

However, despite their expressive power, LLMs exhibit structural limitations. Because their knowledge is stored in fixed model parameters, it reflects the state of data available at training time and cannot be updated dynamically without retraining or fine-tuning. This static knowledge representation may lead to factual inaccuracies, outdated responses, and hallucinated content plausible but unsupported statements generated without grounding in verifiable evidence. These limitations motivate mechanisms that augment parametric models with external, updatable information sources [1].

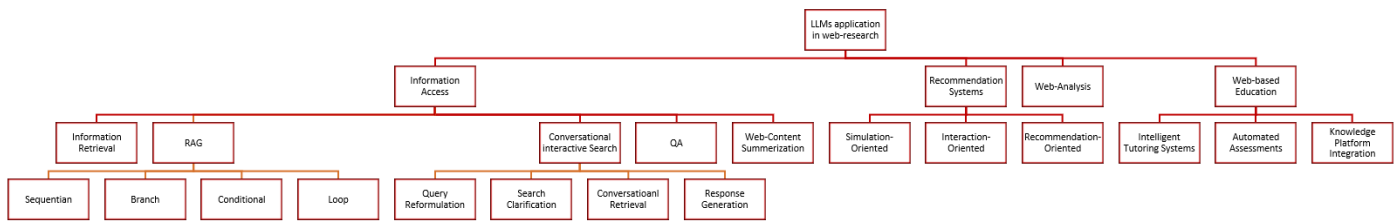

Fig.1 LLM Integration Across Core Web Research Domains

To address the constraints of purely parametric knowledge, modern web-based AI systems increasingly adopt hybrid paradigms that integrate LLMs with external information repositories. In this setting, knowledge is divided into two complementary forms: parametric knowledge, encoded within model weights during pretraining, and non-parametric knowledge, stored in external corpora such as web documents, databases, knowledge graphs, or domain-specific repositories [7].

External retrieval mechanisms enable models to access relevant non-parametric information at inference time, grounding generated outputs in verifiable evidence. This integration establishes a bridge between generative language modeling and traditional information retrieval infrastructures, forming a unified framework in which retrieval and generation operate collaboratively. Such hybrid architectures are particularly significant in web environments, where information is dynamic, large-scale, and continuously evolving. They provide the foundational mechanism through which LLMs interact with the broader web ecosystem and support reliable, context-aware applications across domains [6].

### III. APPLICATIONS OF LLMS IN WEB RESEARCH

#### A. Information Access

Information Retrieval (IR) and AI have historically evolved in close interaction, mutually reinforcing each other's progress. Web-based information serves as a primary data source for training LLMs, while LLMs, in turn, have significantly advanced the capabilities of IR systems. Despite rapid progress, several challenges remain, including the development of more robust models, ethical considerations, bias mitigation, and the continual updating of models to reflect dynamic web content. These issues are discussed across the following subsections, which examine the role of LLMs in different aspects of information access [8].

*1) Information Retrieval*

Information Retrieval plays a foundational role in the modern digital ecosystem. A substantial portion of daily information-seeking activities relies on IR systems, including web search engines such as Google and Microsoft (via Bing), question-answering platforms such as OpenAI's ChatGPT, and content recommendation systems deployed by platforms like YouTube and Amazon [9].

Traditionally, IR methods are categorized into two primary paradigms [9]. The first consists of classical or lexical approaches, including Boolean retrieval, TF-IDF, and BM25 [8]. These techniques retrieve documents from an indexed corpus by computing lexical similarity between a user query and candidate documents. Because they rely on sparse term representations and exact or near-exact token matching, they are commonly referred to as sparse retrieval methods.

While computationally efficient and highly scalable, sparse methods struggle with semantic understanding, synonym resolution, and contextual interpretation, which limits their effectiveness for complex or ambiguous queries. The emergence of pre-trained deep neural networks and Transformer architectures has addressed many of these limitations. Neural and LLM-based techniques enable the construction of dense semantic representations, allowing IR systems to capture contextual meaning rather than relying solely on surface-level lexical overlap. This transition represents not merely incremental improvement, but a paradigm shift in retrieval methodology [8].

More recently, Generative Information Retrieval (GenIR) has further expanded the scope of IR. Broadly, GenIR can be divided into two categories. The first category retains the traditional ranking-based framework but employs generative models for document indexing or representation learning, leading to improved semantic matching and retrieval accuracy [10]. This direction has opened new research areas, including document identifier design, continual learning over dynamic corpora, downstream task adaptation, and multimodal retrieval-generation integration. The second and more transformative direction involves retrieval-integrated response generation, in which systems not only retrieve relevant information but also synthesize coherent, user-centric responses. These systems aim to provide customized, context-aware answers that enhance user engagement and satisfaction. However, generative approaches introduce significant challenges, including hallucinated or irrelevant responses, factual inconsistencies, toxic content generation, and reliance on outdated information. Mitigation strategies include strengthening internal knowledge representations, incorporating external knowledge sources, grounding responses with citations, and improving personalization mechanisms. Despite substantial advancements, ensuring reliability and factual grounding remains a critical open problem in LLM-augmented information retrieval [5], [9].

## 2) RAG

Despite the progress of generative information retrieval, retrieval-integrated generation faces substantial challenges, including prolonged inference time, high computational cost, and risks of hallucination or knowledge conflict. In response, the research community has increasingly adopted RAG as a principled framework for integrating external knowledge into large language models [7]. RAG enables LLMs to access information beyond their parametric memory by dynamically retrieving relevant documents from external corpora during inference [4]. Unlike conventional LLMs, which rely solely on knowledge encoded during pre-training, RAG-based systems extract evidence directly from external databases, thereby reducing dependence on static training data. This mechanism allows models to operate in task-specific or domain-restricted environments such as question-answering systems tailored to particular websites or organizational knowledge bases while maintaining factual grounding. A key advantage of RAG is its flexibility with respect to data types. It can operate over structured (e.g., databases, knowledge graphs), semi-structured (e.g., web pages with metadata), or unstructured text corpora. RAG systems can be categorized along two orthogonal dimensions: architectural interaction patterns and retrieval strategies [4], [6], [7].

### a) RAG Architectures

RAG architectures differ in how retrieval and generation interact.

Sequential RAG follows a retrieve-then-generate pipeline. It is well-suited for factoid questions, single-hop question answering, and news lookup tasks, where queries exhibit clear intent and require a single round of evidence retrieval [4].

Branching RAG processes a query through multiple parallel retrieval–generation branches. This architecture is particularly effective for multi-aspect or multi-sub-question queries that require heterogeneous documents and multi-source evidence aggregation [11].

Conditional RAG introduces a decision mechanism to determine whether retrieval is necessary. It is appropriate for mixed-knowledge scenarios in which some queries can be answered from parametric memory while others require external grounding. This design also reduces unnecessary retrieval cost in latency-sensitive systems and mitigates knowledge conflicts between retrieved documents and the model's internal knowledge [12].

Loop RAG establishes iterative interaction between retriever and generator. Through multi-step retrieval and refinement, it supports complex reasoning, multi-hop question answering, and research-style queries over large unstructured corpora where information needs evolve during generation [12]. All four architectures have been displayed in Fig. 2, Fig. 3 and Fig. 4.

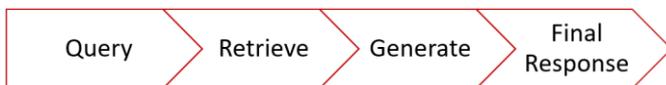

*Fig. 2 Sequential RAG Architecture*

### b) Retrieval Strategies in RAG

Before generation can occur, relevant documents must be selected through an appropriate retrieval strategy. Retrieval strategies are typically categorized into four types as shown in Table 1.

Sparse retrieval, inherited from traditional IR methods such as BM25, relies on lexical matching within high-dimensional sparse term spaces. It is computationally efficient, scalable, and precise for entity-heavy or keyword-specific queries. However, it suffers from vocabulary mismatch and limited semantic generalization, making it less effective for paraphrased or context-dependent queries [8].

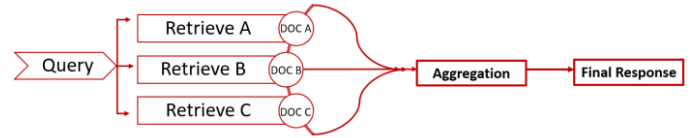

*Fig. 3 Branching RAG Architecture*

Dense retrieval leverages embedding-based similarity via dual-encoder architectures (e.g., DPR-style models). By representing queries and documents as low-dimensional dense vectors, it captures semantic similarity and improves recall in open-domain question answering. Nevertheless, dense models require substantial training, are sensitive to domain shift, and may retrieve semantically similar yet factually irrelevant documents [7].

Hybrid retrieval combines sparse and dense scoring mechanisms to balance lexical precision and semantic recall. While this approach improves robustness across diverse query types, it increases computational cost and system complexity. Iterative retrieval performs multiple rounds of retrieval, refining queries based on intermediate results. This strategy enhances evidence coverage and supports multi-hop reasoning but introduces additional latency and the risk of error propagation [6].

Through the combination of architectural designs and retrieval strategies, RAG significantly enhances the reliability and adaptability of web-based information systems. Importantly, the primary external knowledge source for RAG systems is the web ecosystem itself. In practice, RAG models interact with search engines, knowledge graphs, API-based services, and other model-based tools to acquire up-to-date and domain-specific information, thereby enabling more grounded and context-aware generation [6].

*Table 1. Different Strategies of RAGs*

| Strategy | Strength | Weakness | Best Usecase | Cost |
|---|---|---|---|---|
| Sparse | Fast, Precise | Poor semantics | Entity queries | Low |
| Dense | Semantic recall | Domain shift | Open QA | Medium |
| Hybrib | Balanced | Complex | Mixed queries | High |
| Iterative | Multi-hop | Latency | Research queries | Very High |

### 3) Conversational Active Search and AI Agents

As discussed earlier, the integration of large language models into information retrieval is not solely aimed at improving response accuracy; it fundamentally shifts the paradigm toward user-centered information access. By generating personalized, context-aware, and concise responses tailored to complex multi-turn queries, LLM-based systems significantly enhance usability and interaction quality. Beyond their original purpose, LLMs have been widely leveraged to develop AI agents [13]. While the philosophical definition of an "agent" is broad, in this context we define an AI agent as a computational system capable of interpreting heterogeneous data, making decisions based on contextual signals, and producing actions or generated outputs accordingly. Our focus is specifically on agents that interact with web-based services and search infrastructures. In practical terms, these agents function as conversational interfaces that differ from traditional search engines by engaging in interactive dialogue rather than returning static ranked lists of links [14]. Conversational search represents a major paradigm shift toward interactive information access. Major search providers such as Microsoft's Bing and Google have incorporated LLM-driven conversational capabilities into their systems, aiming to reduce ambiguity and provide more direct, synthesized answers. The distinction between classical search and LLM-, ANN-, or DNN-backed interactive search becomes evident through four key stages: query reformulation, search clarification, conversational retrieval, and response generation [15].

*a) Query reformulation*

Query reformulation restructures user input to enhance multi-turn conversational understanding. This process is essential for maintaining coherence across dialogue turns and is typically implemented through three techniques: query expansion, query rewriting, and query decomposition. Within RAG-based architectures, particularly sequential and iterative (loop-based) frameworks, retrieval feedback can implicitly improve reformulation quality by aligning query representations with downstream retrieval and generation objectives. This feedback loop enhances the semantic alignment between user intent and retrieved evidence [7].

*b) Search Clarification*

Conversational agents improve retrieval precision by engaging users in clarification dialogue when ambiguity is detected. Instead of providing an immediate answer, the system may request additional details to refine user intent. While clarification enhances personalization and relevance, detecting when clarification is necessary is non-trivial. Excessive clarification increases latency and user friction, whereas insufficient clarification risks misunderstanding. Conditional variants of RAG-style systems introduce decision mechanisms that determine whether retrieval or additional interaction is required, thereby improving efficiency and cost control [15][14].

*c) Conversational Retreival*

After reformulation and clarification, the system retrieves relevant information from external knowledge sources. Conversational retrieval differs from traditional ad hoc retrieval in that it must account for extended dialogue history and evolving user intent. A naïve approach rewrites the latest query and applies standard retrieval. However, this strategy degrades as conversational context grows longer and more complex. A more advanced approach encodes the entire dialogue session into a dense representation, enabling retrieval directly from contextual embeddings rather than relying solely on explicit rewriting. This session-level encoding improves optimization through ranking signals and enhances alignment between retrieval and downstream generation. However, encoding long conversational histories increases computational cost and model complexity. Following retrieval, context-aware re-ranking refines document ordering to ensure that the generation module receives the most relevant and coherent evidence, maintaining consistency across the conversational workflow [15], [16].

*d) Response Generation*

Unlike traditional search engines that provide ranked lists, conversational search agents synthesize retrieved information into explicit, dialogue-form responses. Outputs may include concise direct answers, summaries, or structured formats such as tables, depending on user needs. This stage presents several challenges: Aligning response format with user intent; Preventing hallucinations during integration of retrieved and internal knowledge managing conflicts between dialogue context, external sources, and parametric model knowledge handling long-context generation; Providing transparent citations for verifiability Architectures based on RAG for Knowledge-Intensive NLP Tasks mitigate many of these issues by grounding generation in external evidence, reducing hallucination risk, and improving factual consistency across conflicting sources [16].

Modern LLM-based conversational frameworks fundamentally transform web search from static retrieval to interactive, adaptive information access. These systems are powered by web-scale data both during training and inference, creating a reciprocal relationship: web content fuels model capability, while LLM-based agents redefine how users interact with web-based information systems [14], [16].

*4) QA Web System*

Although web-based Question Answering (QA) systems substantially overlap with conversational search in both functionality and limitations and can be regarded as a component within it, their central role in web research justifies a dedicated discussion. Question Answering is a specific task within natural language processing and information retrieval that aims to produce a precise, contextually relevant answer to a user's query, rather than returning a ranked list of documents. Traditional QA systems were typically pipeline-based and modular, often requiring large annotated datasets, exhibiting limited generalization across domains, and relying on static retrieval–ranking architectures. The emergence of large language models, particularly when combined with RAG has significantly mitigated these limitations by enabling dynamic reasoning, improved contextual understanding, and integration of up-to-date external knowledge. The four-stage architecture of LLM-based QA agents: planning, question understanding, retrieval, and answer generation, closely mirrors the classical conversational search workflow. However, while conversational search describes the interactional pipeline from the user's perspective, agent-based QA formalizes the internal cognitive and decision-making mechanisms that govern retrieval, evidence selection, and response synthesis[17].

Despite these advances, web QA systems continue to face key challenges, including hallucination, knowledge conflicts between parametric and retrieved information, long-context reasoning, citation generation, and robustness to multi-hop queries. These challenges motivate ongoing research into more reliable, interpretable, and adaptive QA-web frameworks capable of sustained and trustworthy information access [13][16].

*5) Web-Content Summarization*

Among the applications of LLMs, summarization is a key feature, leveraging their language understanding and generative capabilities. When combined with RAG, LLMs can produce concise, contextually grounded summaries of web content, supporting tasks such as condensing online videos, aggregating website comments, and providing accurate, evidence-based information for web QA systems [18].

B. *Web-Based Recommendation and Personalization*

Alongside information access and retrieval, recommender systems (RS) constitute one of the central research areas in web science and have been significantly influenced by the recent advancements in LLMs. Recommender systems underpin personalization across modern web platforms and play a crucial role in enhancing user engagement, satisfaction, and retention. In this section, we first provide an overview of recommender systems and their traditional methodologies. We then analyze how LLMs interact with and enhance these systems, and finally discuss the emerging limitations and open challenges introduced by LLM-driven personalization [19].

*1) How do Recommendation System Work?*

Recommender systems are designed to predict user preferences and choices across various web platforms, from e-commerce websites to online entertainment services and social media. Traditional recommender systems are generally based on three major approaches: collaborative filtering, content-based filtering, and hybrid methods. Collaborative filtering assumes that a group of users share similar preferences and therefore recommends items favored by users within the same similarity group. In another strategy, the model groups items based on interaction patterns and recommends items similar to those previously selected by a user [19].

Since this process typically relies on user–item rating matrices, sparsity in these matrices can lead to both computational challenges and reduced reliability of predictions. In contrast, content-based filtering utilizes attributes of items and user profiles to generate recommendations. This approach addresses some limitations of collaborative filtering, particularly dependence on other users' data. However, it still faces challenges due to incomplete or low-quality item attribute information and limited semantic precision, often leading to over-specialization toward certain attributes. Finally, the hybrid approach combines both collaborative filtering and content-based methods, leveraging user interaction data alongside item attributes for prediction. Although this approach resolves some limitations of the individual methods, sparsity and scalability remain major computational challenges for recommender systems. Moreover, issues such as the cold-start problem, static modeling of user preferences, and lack of interpretability continue to affect their robustness and overall effectiveness [19].

*2) LLM-Enhanced Recommendation Systems*

With the emergence of LLMs, recommender systems remain in a relatively early stage of integration; however, numerous approaches have been proposed to incorporate LLMs into recommendation pipelines. These approaches can generally be categorized into three paradigms as summarized in Table 2. [20].

*a) Simulation-Oriented*

One prominent direction focuses on modeling user intent based on user activities such as clicks, queries, and other interaction records. This can be implemented using various models, ranging from classical machine learning methods to deep learning algorithms. By leveraging such interaction data, systems construct user models and simulate user preferences through hypothetical or synthetic queries. More comprehensive preference analysis, behavioral tracking, and pattern modeling can lead to more accurate predictions and improved evaluation of recommender systems [20].

*b) Interaction-Oriented*

As previously discussed in conversational agents, LLMs in this paradigm act as interactive agents that extract user preferences through dialogue and ultimately provide interpretable recommendations using natural language processing. These systems generate human-understandable justifications explaining why a particular item is suggested, which enhances transparency and user engagement. However, challenges such as hallucination risks and high computational costs remain significant concerns for these models [19].

*c) Recommendation-Oriented*

In this paradigm, the LLM operates more independently as a reasoning agent that directly generates recommendation decisions. Two major approaches are particularly relevant. The first is cross-domain recommendation, where user preferences in one domain (e.g., books) are inferred from knowledge in another domain (e.g., movies). This is facilitated by the embedding and semantic representation capabilities of LLMs, which enable alignment across feature spaces and support semantic transfer [21].

*Table 2. Categorizaiton of Recommendation Systems paradigms*

| Concept | Recommendation-Oriented | Interaction-Oriented | Simulation-Oriented |
|---|---|---|---|
| User intent modeling | ✓ (direct decision) | ✓ (dialogue refinement) | ✓ (synthetic intent) |
| Review-based Recommendation | ✓ | | |
| Conversational Recommendation System | | ✓ | ✓ |
| Cross-domain Recommendation | ✓ | | |
| Few-shot Personalization | ✓ | | |

By aggregating user information across domains, this approach can help mitigate the cold-start problem, although some domain-specific attributes may be lost during transfer. The second approach is few-shot personalization, which relies on limited available user data to generate personalized predictions. Rather than requiring extensive retraining, the model conditions its outputs on minimal interaction signals. While this method may produce relatively shallow personalization, it is particularly suitable for cold-start scenarios where only sparse user information is available [22].

*3) LLMs Architetcures for Recommendation Systems*

Each enhancement paradigm requires a corresponding LLM architecture. These architectures can be categorized according to their role within the recommendation pipeline, learning strategy, and level of integration. Here, we briefly outline their structural patterns in recommender systems [22][23].

*a) LLMs as Feature Generator*

In this architecture, the LLM produces embeddings or semantic features extracted from textual sources such as user reviews, item descriptions, queries, or social connections. The resulting representations are then incorporated into downstream models, including collaborative filtering, graph neural networks (GNNs), or hybrid recommenders. This approach enhances the semantic understanding of user–item relationships without requiring retraining of the underlying recommendation model. Such architectures can leverage Branching-RAG for parallel feature retrieval and Sequential-RAG for multi-step feature extraction [24].

*b) LLMs as Reasoning Layer*

Here, LLMs directly perform ranking, scoring, and reasoning over user and item data. By utilizing Loop, Conditional, and Sequential-RAG variants, the model can iteratively refine rankings, adapt reasoning based on user context, and conduct multi-step evaluation of candidate items. While this architecture provides stronger reasoning capabilities, it incurs significant computational cost and remains vulnerable to hallucination [25].

*c) LLMs as Dialogue Interface*

In this setting, LLM-based agents act as conversational interfaces that improve user engagement, transparency, and personalization. Through multi-turn interaction, the system can dynamically elicit user preferences and provide interpretable recommendations. This architecture is particularly aligned with interaction-oriented paradigms [20].

*d) Hybrid Architectures*

Hybrid architectures combine LLM-based components with traditional recommender models. Examples include integrating LLM-generated embeddings with collaborative filtering, combining LLM reasoning with knowledge graphs, or orchestrating multiple RAG strategies within a unified pipeline. While highly flexible and powerful, such architectures introduce integration complexity and increased inference cost [26].

*4) Limitations and Challenges of LLMs for Recommendation Systems*

We have already discussed general challenges of LLMs throughout previous sections, such as hallucination and high computational costs. However, several issues are particularly critical in the context of recommender systems. The first is over-personalization, where overly deep or narrow feature extraction may excessively specialize recommendations, reducing diversity and limiting exposure to alternative items. Another persistent challenge is the cold-start problem. Although LLM-based approaches have alleviated this issue to some extent particularly through semantic reasoning and cross-domain knowledge transfer it remains unresolved in scenarios with extremely limited user data. Finally, privacy concerns and ethical issues related to the collection, storage, and processing of user data remain significant. The use of large-scale textual interaction logs and personal behavioral data raises risks regarding data leakage, user profiling, and regulatory compliance [24], [26].

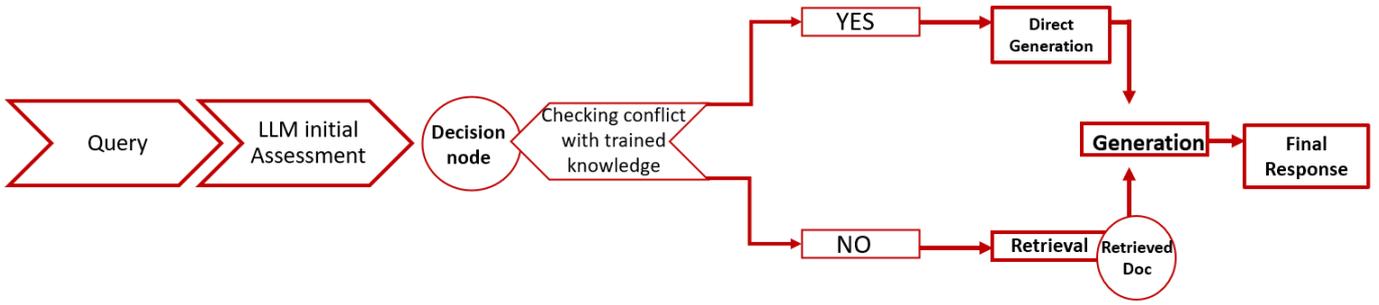

Fig. 3 Conditional RAG Architecture

## C. Web-Analytics and Mining

The rapid expansion of the Internet has generated large-scale, heterogeneous data, enabling analytics across a wide range of domains. Web mining encompasses diverse applications, including log analysis, sentiment analysis, event detection, trend discovery, and content classification, serving purposes from commercial optimization to political monitoring. Core tasks in web analytics and mining include web content classification, opinion mining, event detection (e.g., monitoring online document streams), log analysis for discovering user navigation patterns (a key data source for recommender systems), web information extraction, trend detection, and web clustering [27].

In the pre–deep learning era, web mining was primarily a statistical and algorithmic discipline grounded in classical IR techniques and rule-based systems [28]. These approaches relied heavily on feature engineering, probabilistic modeling, and manually designed heuristics, with limited representation learning. The early 2010s marked a turning point with the broader adoption of machine learning techniques [29]. For instance, Kaya et al.[30] applied Naïve Bayes, Support Vector Machines (SVM), and n-gram features to sentiment classification tasks, reflecting the dominant paradigm of supervised linear models. Subsequently, deep learning architectures further advanced performance, particularly in sentiment detection and text classification, by enabling automated feature extraction.

The emergence of LLMs in the late 2010s introduced a new paradigm in web analytics. Leveraging large-scale pretraining, LLMs demonstrate strong capabilities in text generation, involving heterogeneous web data sources [3].

However, while LLMs substantially advance the analytical capacity of web mining systems, they do not eliminate classical challenges. Issues such as data quality, generalization under distribution shift, and computational cost remain central concerns. Moreover, LLM-based analytics introduce additional risks, including bias amplification, fairness violations, limited interpretability, and political or societal impact. Since web analytics outputs often serve as foundational inputs for downstream decision-making systems, deviations or distortions at this stage may propagate across broader web-based applications. Therefore, responsible deployment and evaluation frameworks are essential when integrating LLMs into web mining pipelines. contextual language understanding, machine translation, and knowledge integration. Their semantic modeling and contextual awareness enable more flexible and unified handling of web mining tasks, often replacing multi-stage pipelines with prompt-based or generative inference frameworks. Furthermore, LLMs facilitate cross-domain transfer and multimodal integration, which accelerates tasks [1].

## D. Web-Based Education and Knowledge Platforms

The spread of the internet access and some global events such as the COVID-19 pandemic led people to use the internet's capabilities for education and online courses [31]. Although this is not a novel phenomenon, its popularity has grown in recent years and has also leveraged the abilities of LLMs [32].

Web-based education using LLMs can be evaluated in three categories: intelligent tutoring systems, automated assessments, and knowledge platform integration. These categories are essentially wise usages of what we have already discussed, and due to the importance of education in the age of AI, we briefly review these applications.

Adaptive tutoring can be interpreted as a learning recommendation system, where, by recognizing a student's characteristics, appropriate teaching materials are offered [31]. In this process, all RAG architectures play their roles: Sequential RAG retrieves relevant lesson material; Branching RAG chooses different explanation strategies; Conditional RAG selects the difficulty level based on the student's level; and Loop RAG iterates until understanding is achieved. Another perspective on adaptive tutoring is the use of conversational active search as a clarification tool for students, providing them with refined explanations [16].

LLMs in education can also benefit instructors in essay scoring, question generation, and feedback generation [33], [34]. Framing teaching syllabi can be another advantage in the teaching process. Loop RAG and Conditional RAG are particularly suitable for these goals, since these tasks require reasoning and multi-step evaluation prompting. Finally, we have knowledge platforms using LLMs. Well-known platforms such as edX and Udemy use LLMs for primitive tasks such as recommendation, search, and question answering, and organizations such as Khan Academy use AI-based tutors and LLM-based examiners.

Surely, the more interaction there is with humans, the more concerns arise. These concerns are not only technical issues but also important moral considerations that must be addressed carefully and are challenging for the use of LLMs in educational systems. Using AI can threaten academic integrity, increase over-dependency, and affect reliability [35]. It is no secret that today a wide range of academia—from university professors to students, and from researchers to reviewers—use AI as an assistant to speed up their work. It is an inevitable issue that requires accurate regulation, moral codes, and technical platforms such as AI-generated content detection tools to prevent wrongdoing that could harm the scientific community.

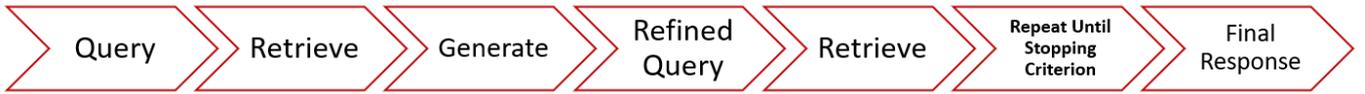

Fig. 4 Loop RAG Architecture

## IV. FUTURE DIRECTIONS

Although LLMs have significantly advanced web-based systems, several challenges remain open. Ensuring factual reliability and grounding is critical, particularly in retrieval-augmented and conversational settings where hallucination and knowledge conflicts persist. More robust verification and citation-aware generation mechanisms are needed. Scalability and efficiency also require attention, as web-scale deployment demands low-latency and cost-effective inference. Advances in adaptive retrieval, model compression, and optimized architectures will be essential. In addition, evaluation standards must evolve to measure generative quality, personalization effectiveness, robustness, and fairness in unified frameworks. Finally, privacy and ethical governance remain central concerns, especially in recommendation and educational systems that rely on sensitive user data. Addressing these issues will shape the next generation of reliable and responsible LLM-driven web infrastructures.

## V. CONCLUSION

Large Language Models have introduced a structural shift in web research by integrating retrieval, generation, reasoning, and interaction within unified architectures as shown in Table 3. Across domains such as information retrieval, recommender systems, conversational agents, web analytics, and education platforms, LLMs function as a common semantic and generative layer. This survey provided a cross-domain perspective on architectural patterns, integration strategies, and shared challenges. While current systems demonstrate substantial capabilities, long-term progress depends on improving reliability, scalability, evaluation, and ethical deployment. The continued evolution of web-based intelligence will be closely tied to principled integration of LLMs within dynamic information ecosystems.

*Table 3. LLMs effects in Web Research*

| Aspect | Traditional Systems | LLM-Based Systems |
|---|---|---|
| Architecture | Modular pipeline | Integrated generative-retrieval |
| Knowlege | Explicit index | Parametric trained and external |
| Interaction | Static query | Conversational |
| Personalization | Feature-based | Semantic and reasoning |